# High Performance Photocathodes based on Molecular Beam Epitaxy Deposition for Next Generation Photo Detectors and Light Sources

Junqi Xie[a], Marcel Demarteau[a], Robert Wagner[a]

[a]High Energy Physics Division, Argonne National Laboratory, Lemont, USA 60439

**Abstract**

*The development of high-performance photocathodes is a key challenge for future accelerator and particle physics applications. In this paper photocathode growth through molecular beam epitaxy is introduced as a promising technique to obtain robust, highly efficient alkali-antimonide based photocathodes. Recent research shows that the quantum efficiency of photocathodes can be significantly enhanced through control of the photocathode crystallinity. Molecular beam epitaxy allows for cost-effective growth of large-area photocathodes with excellent control of the stoichiometry and crystallinity, making photocathodes with peak quantum efficiencies exceeding 35% routine.*

## 1. Introduction

In high energy physics experiments, photodetectors are used in a wide array of particle detection methods, such as time of flight (TOF) measurements, calorimetry and the detection of Cherenkov or scintillation light [1-2]. Proposed or upgraded detectors in particle and nuclear physics call for next generation photodetectors with low cost, large area, high quantum efficiency (QE), high gain, low dark current and fast response. Meanwhile, the next generation free-electron laser (FEL) x-ray light sources require electron injectors with high repetition rate, high brightness, low emittance and long lifetime. Both areas of science need robust photocathodes with improved performance. Their development is a key challenge and a high priority in the accelerator community [3-6].

Many different kinds of photocathode materials have been used in practical applications. Most of the cathodes are made of compound semiconductors, such as GaAs:Cs, $Cs_2Te$, $Cs_3Sb$ and $K_2CsSb$, consisting of alkali metals with a low work function. GaAs:Cs photocathodes [7-8] have high QE (~20%) over a broad range of wavelengths. However, GaAs:Cs photocathodes suffer from saturation at high pulse charge, relatively long pulse length, extreme sensitivity to surface contamination (vacuum requirements of $10^{-11}$ Torr or better) and are sensitive to ion-induced lattice damage. $Cs_2Te$ photocathodes [9] exhibit relatively low QE (~3% at UV). Alkali antimonide photocathodes [10-11] have modest QE (20% at 380 nm and >5% at 532 nm), exhibit very fast time response (at the pico-second level) due to its positive electron affinity surface condition, and only require $10^{-8}$ Torr vacuum, which gives these devices a long lifetime. This level of vacuum requirement can be easily fulfilled with current high vacuum technology. Alkali antimonide photocathodes are thus considered a promising candidate photocathode as a path towards higher QE (>40% at UV and >15% at 532 nm).

## 2. Development History of Alkali Photocathodes

The development of the alkali-antimonide photocathode was relatively slow over the past several decades, due to a poor understanding of the material's microscopic properties. The development mainly occurred within companies for a given application. For example, the peak QE was very low (<1%) when the alkali-antimonide photocathode was discovered in the 1930s. It was soon improved to a few percent and then to 25% in the mid 1960s by RCA, USA. Recently, Hamamatsu has reported super-bialkali photocathodes with a QE as high as 43% at 380 nm by improving the photocathode crystallinity [12]. To the best of our understanding, all of the current alkali-antimonide photocathodes were obtained through sequential diffusion of alkali (Na, K, Cs) vapors on a pre-deposited antimony

(Sb) layer to form alkali-antimonide compounds. Sb/Pt beads on a Mo/W wire and various alkali compound dispensers are used as element sources. During the production, these sources are electrically heated to release the elemental vapors. This method is quite cost-effective, though a major drawback is the fact that the diffusion process cannot be precisely controlled to form stoichiometric compounds with the correct crystalline structure to build up designed photocathode structures. Previous improvements were mainly obtained through an empirical process, paying considerable attention to production details.

Recently, a series of real-time in-situ x-ray studies [13-15] indicate that K-Cs-Sb photocathodes following the traditional diffusion methods are mixtures with different stoichiometry and the resulting structures are either amorphous or polycrystalline. The polycrystalline K-Cs-Sb photocathode with stoichiometry of K:Cs:Sb=2:1:1 exhibits a much higher QE than other configurations. The correct stoichiometry improves the photocathode crystallinity, decreases the defect scattering, and increases the electron mean free path, thereby achieving a higher QE and lower dark current. A stoichiometric crystalline $K_2CsSb$ photocathode with modified doping profile is expected to achieve even higher QE and lower dark current.

## 3. Photocathode Deposition based on the MBE Technique

The Molecular Beam Epitaxy (MBE) technique, which has been developed in the thin film community, perfectly fits the requirements for high performance photocathode deposition:
1. Precise control of the elemental beam flux for the stoichiometric compound;
2. Well developed technology for crystalline material deposition;
3. Ability to fabricate device structures with designed doping profiles;
4. Easily adopted for use in large-area planar photodetectors.

By applying the MBE technique to the development of photocathodes, stoichiometric crystalline alkali-antimonide photocathodes can be grown. Due to the right stoichiometry K:Cs:Sb=2:1:1 and perfect crystalline structure, the photocathodes are expected to exhibit much higher QE and lower dark current. At the same time a doping profile across the photocathode structure can be designed to further enhance the performance of the alkali photocathode [16]. The doping profile is designed to have low p doping near the substrate, high p doping near the vacuum surface and possibly an atomic layer of n-doping for maximum free electron injection. We anticipate such a photocathode to exhibit a peak QE in excess of 30%, possibly achieving 40% after structure optimization. The dark count rate is expected to be lower than current commercial photocathodes.

Furthermore, the MBE photocathode growth method that we propose to develop can be easily enlarged and used in novel, large-area planar photodetectors. These detectors are currently being developed at Argonne in collaboration with the University of Chicago and Space Science Laboratory at Berkeley.

## 4. Summary

A new method to grow high performance (high QE, high robustness, low dark current and fast response) photocathodes for next generation photodetectors and light sources is proposed. The well-established MBE deposition technique is used to obtain crystalline photocathode structures with the correct stoichiometry for high performance. The doping profile across the photocathode structure can also be designed and optimized to further enhance the photocathode performance. These alkali-antimonide photocathodes are expected to benefit a wide range of applications besides use in photodetectors and accelerators.

# 5. References

[1] Alimonti G. et al. "Science and technology of Borexino: a real-time detector for low energy solar neutrinos", *Astrop. Phys.* 16:205-234, **2002**.

[2] Eguchi K. et al. "First Results from KamLAND: Evidence for Reactor Antineutrino Disappearance", *Phys. Rev. Lett.* 90:021802, **2003**.

[3] Basic Energy Science Advisory Committee, U.S. Department of Energy, "Report of the Basic Energy Sciences Workshop on Compact Light Sources", May **2010,** http://science.energy.gov/~/media/bes/pdf/reports/files/CLS.pdf

[4] Basic Energy Science Advisory Committee, U.S. Department of Energy, "Next Generation Photon Sources for Grand Challenges in Science and Energy", May **2009,** http://science.energy.gov/~/media/bes/pdf/reports/files/ngps_rpt.pdf

[5] "R&D for a soft X-ray free electron laser facility", June **2009,** http://www.escholarship.org/uc/item/8rc959p8

[6] "Science and Technology of Future Light Sources", December **2008,** http://www-ssrl.slac.stanford.edu/content/sites/default/files/documents/science-technology-of-future-light-sources.pdf, ANL-08/39, BNL-81895-2008, LBNL-1090E-2009, SLAC-R-917

[7] C. Herrnandez-Garcia *et al.*, "A high average current DC GaAs photocathode gun for ERLs and FELs", *Proceedings of the 2005 Particle Accelerator Conference*, Knoxville, TN, May **2005**

[8] L.B. Jones *et al.*, "Photocathode preparation system for the ALICE Photoinjector", *AIP Conf. Proc.* 1149, 1089 **2009**

[9] S. Kong *et al.*, "Photocathodes for free electron lasers"*, Nucl. Instrum. Methods Phys. Res. A* 358*,* 272 *(***1995***)*

[10] J. L. McCarter *et al.*, "Performance study of $K_2CsSb$ photocathode inside a DC high voltage gun", *Proceedings of 2011 Particle Accelerator Conference*, New York, NY, USA, March **2011**

[11] Hamamatsu PMT Handbook, http://hamamatsu.com

[12] K. Nakamura et al., "Latest bialkali photocathode with ultra-high sensitivity", Nucl. Instr. and Meth. A 623 (2010) 276

[13] M. Ruiz-Oses *et al., "*In-situ characterization of bi-alkali antimonide photocathodes for high brightness accelerators", *Appl. Phys. Lett. Mater.* **2013**

[14] J. Xie, *et al.,* "Real Time Evolution of Antimony Deposition for High Performance Alkali Photocathode Development", *Proc. SPIE,* 8847, **2013**.

[15] S. Schubert *et al.*, "Surface characterization of bi-alkali antimonide photocathodes for high brightness accelerators. *Appl. Phys. Lett. Mater.*, **2013**

[16] A. H. Sommer, "Photoemissive Materials", Robert E. Krieger Publishing Company, 1980.